\documentclass[useAMS,usenatbib]{mn2e}
\usepackage[a4paper,body={17cm,23cm},foot=2cm,head=15mm]{geometry}
\usepackage{graphicx}
\usepackage{times}
\usepackage{booktabs}
\usepackage{longtable}
\usepackage{amsmath}
\usepackage{amsfonts}
\usepackage{amssymb}
\usepackage{subfigure}
\usepackage{rotate}
\usepackage{wasysym}
\usepackage{verbatim}
\usepackage{upgreek}

\footnotesize
\newdimen\digitwidth    
\setbox0=\hbox{\rm0}
\digitwidth=\wd0
\catcode`!=\active
\def!{\kern\digitwidth}
\normalsize

\title[HTRU-North I]{The Northern High Time Resolution Universe Pulsar Survey I:\\Setup and initial discoveries}

\author[E.~D.~Barr et al.]{\parbox{\textwidth}{
E.~D.~Barr,$^{1,2,3}$\thanks{E-mail: ebarr@swin.edu.au}
D.~J.~Champion,$^{1}$
M.~Kramer,$^{4,1}$
R.~P.~Eatough,$^{1}$
P.~C.~C.~Freire,$^{1}$
R.~Karuppusamy,$^{1,4}$
K.~J.~Lee,$^{1}$
J.~P.~W.~Verbiest,$^{1}$
C.~G.~Bassa,$^{4}$
A.~G.~Lyne,$^4$ 
B.~Stappers,$^4$
D.~R.~Lorimer,$^5$
B.~Klein$^{1,6}$}
\vspace{0.4cm}
\\
\parbox{\textwidth}{
$^{1}$ Max-Planck-Institut f\"ur Radioastronomie, Auf dem H\"ugel 69, 53121 Bonn, Germany\\
$^{2}$ Centre for Astrophysics and Supercomputing, Swinburne University of Technology, Mail H30, PO Box 218, Hawthorn, VIC 3122, Australia\\ 
$^{3}$ Australian Research Council Centre of Excellence for All-Sky Astrophysics (CAASTRO), Mail H30, PO Box 218, Hawthorn, VIC 3122, Australia\\
$^{4}$ Jodrell Bank Centre for Astrophysics, School of Physics and Astronomy, The University of Manchester, M13 9PL, UK\\
$^{5}$ Department of Physics, West Virginia University, White Hall, Morgantown, WV 26506, USA\\
$^{6}$ University of Applied Sciences Bonn-Rhein-Sieg, Grantham-Allee 20, 53757 Sankt Augustin, Germany\\}}

\date{Received: --Accepted:}

\begin{document}
\maketitle
\newcommand{\setthebls}{
}
\setthebls
\begin{abstract}
We report on the setup and initial discoveries of the Northern High
Time Resolution Universe survey for pulsars and fast transients, the first major pulsar survey conducted with the 100-m Effelsberg radio telescope and the first in 20
years to observe the whole northern sky at high radio frequencies. Using a newly developed
7-beam receiver system combined with a state-of-the-art polyphase
filterbank, we record an effective bandwidth of 240 MHz in 410 channels centred on
1.36 GHz with a time resolution of 54 $\mu$s. Such fine time and
frequency resolution increases our sensitivity to millisecond pulsars
and fast transients, especially deep inside the Galaxy, where previous
surveys have been limited due to intra-channel 
dispersive smearing. To optimise observing time, the survey is split into three
integration regimes dependent on Galactic latitude, with 1500-s, 180-s
and 90-s integrations for latitude ranges $|b|<3.5^{\circ}$,
$|b|<15^{\circ}$ and $|b|>15^{\circ}$, respectively. The survey has so
far resulted in the discovery of 15 radio pulsars, including a pulsar
with a characteristic age of $\sim18$ kyr, {PSR~J2004+3429}, and
a highly eccentric, binary millisecond pulsar,
{PSR~J1946+3417}. All newly discovered pulsars are timed using the
76-m Lovell radio telescope at the Jodrell Bank Observatory and the Effelsberg radio telescope. We
present timing solutions for all newly discovered pulsars and discuss
potential supernova remnant associations for {PSR~J2004+3429}.
\end{abstract}

\section{Introduction}

It could be argued that no other astrophysical object has the ability
to provide insight into as many fields of physics and astrophysics as
the pulsar. The extreme conditions found in and around these objects
make them unique natural laboratories for the study of subjects such
as the equation of state of supra-nuclear matter
\citep{Demorest2010,Antoniadis2013}, the behaviour of gravity in the strong-field
regime \citep{Kramer2006,Freire2012a}, the formation and evolution of binary
systems \citep{Stairs2004} and the existence and properties of
gravitational waves \citep{Hellings1983}. Therefore the discovery of new
pulsar systems, through targeted or blind surveys, holds great
scientific potential.

Pulsar surveys will in general always increase our understanding of
the underlying source distribution and its properties, but it is the
potential for the detection of rare and exciting systems, such as a
hypothesised pulsar-black hole system \citep{Narayan1991}, that is the major
driving force behind modern-day pulsar surveys. Several examples of
such exciting discoveries can be found in surveys conducted within the
last two decades. These include the discovery of the so-called `Double pulsar',
PSR~J0737$-$3039A/B \citep{Burgay2003,Lyne2004}, which consists of two pulsars
orbiting each other in a highly relativistic binary system; of
{PSR~J1903+0327} \citep{Champion2008}, a rapidly rotating pulsar in a
highly eccentric orbit, which has shed light on the evolution of hierarchical triple systems \citep{Freire2011}; of the `Diamond-planet pulsar', {PSR~J1719$-$1438} \citep{Bailes2011}, with
its Jupiter-mass CO white dwarf companion; and most recently of J2222$-$0137 \citep{Boyles2012}, a rapidly rotating pulsar with a massive companion, whose proximity to the Earth ($\sim300$ pc) makes it an exciting system, both for the measurement of post-Keplerian parameters and for the multiwavelength study of pulsar emission physics. 

While targeted pulsar surveys, such as those that observe globular
clusters \citep[e.g.][]{Ransom2005a} or $\gamma$-ray point sources
\citep[e.g.][]{Keith2011,Barr2012} tend to have a high discovery
rate, they cannot produce an unbiased sample of the underlying population. To achieve a more complete picture of
the true population distribution and the exotic systems it may contain,
we must perform all-sky surveys.

In the past, blind surveys have been successful in detecting many new
and exciting pulsar systems. Good examples can be found in the many
surveys \citep[e.g.][]{Manchester2001,Burgay2006a,Edwards2001}
conducted using the 20-cm multi-beam receiver system of the Parkes
Radio Telescope over the last 10 years. These surveys have been
remarkably successful, not only discovering almost 60\% of all known
pulsars, but also some unique and fascinating objects. As well as the
discovery of the aforementioned Double pulsar, these surveys have
discovered six of the ten known double neutron star systems (DNS), pulsars with massive stellar companions \citep{Johnston1992,Stairs2001} and the pulsar with the largest glitch
\citep{Manchester2011}. Furthermore, reprocessing of these data has led to the discovery of Rotating Radio Transients \citep[RRATs;][]{McLaughlin2006}, a new class of pulsars that display bursty radio emission on varying timescales \citep[see][for a recent review]{Keane2011a}.

The original Parkes multibeam surveys used a 96$\times$3-MHz channel analogue filterbank with a 250-$\mu$s sampling time. This relatively coarse frequency and time resolution resulted in a reduced searchable volume for narrow-pulse-width transients and millisecond pulsars (MSPs) due to dispersive smearing within individual channels (see Section \ref{htru:dedispersion}). These limitations were compounded by a 1-bit digitisation scheme employed by the analogue filterbank. The limited dynamic range of 1-bit digitisation, although less sensitive to radio-frequency interference (RFI), acts to decrease the signal-to-noise (S/N) ratio of any pulsed or transient signal by $\sim$20\% \citep{Kouwenhoven2001}. While these surveys were state-of-the-art at their conception, affordable technology now exists to significantly improve on them. Recent advancements in field-programmable gate array technology and data transfer and storage techniques, have allowed for the use of
tunable polyphase filterbanks capable of providing large numbers of
high-resolution frequency channels and high sampling rates. These advances, combined with the development of a state-of-the-art 1.36-GHz multi-beam receiver for the 100-m Effelsberg telescope, have led to the commencement of the Northern High Time Resolution
Universe (HTRU-North) survey, the motivation behind which is the discovery of exotic and extreme pulsar systems, and the characterisation of the transient sky down to timescales of a
few tens of microseconds.

The HTRU-North survey is half of a full-sky survey being undertaken using both the 100-m Effelsberg radio telescope in the northern hemisphere and the 64-m Parkes radio telescope in the southern hemisphere. Together, these surveys aim to achieve complete sky coverage at varying integration depths, with the deepest integrations along the Galactic plane (see Section \ref{htru:strategy}). 

The southern hemisphere half of the survey \citep[High Time Resolution Universe;][]{Keith2010} has been
underway for several years, resulting in the discovery of more than
100 new pulsars, several of which are high-dispersion-measure MSPs
\citep{Bates2011,Keith2012} that were most likely undetected in previous surveys due to intra-channel dispersive smearing.

While Parkes has had a long, successful history with pulsar searches, the Effelsberg telescope has seen limited use as a pulsar survey instrument. This work marks the start of a new era of blind pulsar surveys with the Effelsberg telescope, building on the successes of \citet{Lorimer1999} and \citet{Barr2012} to clearly establish Effelsberg as a powerful instrument for pulsar searches.

In Section \ref{htru:strategy} we describe the observing strategy used
in the HTRU-North survey.  In Section \ref{htru:instrumentation} we
describe the frontend and backend systems used at the Effelsberg
radio telescope. In Section \ref{htru:sensitivity} we present
analytical and empirical estimates of the survey sensitivity. In
Section \ref{htru:simulations} we consider the expected pulsar yield as
determined through Monte Carlo simulations of the Galactic pulsar
population. In Section \ref{htru:processing} we describe the data
processing pipeline from acquisition to candidate selection. In Section
\ref{htru:discoveries} we present the timing solutions for all newly discovered pulsars and discuss potential supernova remnant associations for PSR J2004+3429. In Section \ref{htru:conclusion} we present our conclusions.

\section{Survey strategy}
\label{htru:strategy}

The HTRU-North survey will be comprised of more than 1.5 million observed
positions on the sky. To optimise the usage of our observing time, the HTRU-North survey is split into three complementary parts based on Galactic latitude.

The \textbf{high-latitude} section covers the sky at Galactic
latitudes of $|b| > 15^{\circ}$ with short integrations of 90 s. The
majority of sky covered by the high-latitude section has remained
unsurveyed for more than 20 years, and so with the technical advances implemented
in the HTRU-North survey we expect to discover many bright pulsars, which do not require long integration times to detect, and
both Galactic and extragalactic transients (see Section \ref{htru:processing}). The near-isotropic distribution of  MSPs expected to be discovered in this region will be of great use to current and future pulsar timing arrays for gravitational-wave detection \citep{Foster1990}.

The \textbf{mid-latitude} section covers Galactic latitudes of $|b| <                                                                                                                
15^{\circ}$ with 180-s integrations. This section of the survey
probes the regions of the Galaxy most likely to contain undiscovered bright MSPs (again, vital for pulsar timing arrays). In the mid-latitude section, we also perform a shallow sweep of the Galactic plane. These observations are expected
to discover any bright, longer-period pulsars.

Finally, the \textbf{low-latitude} section covers Galactic latitudes of $|b| < 3.5^{\circ}$ with long integrations of 1500 s. These long integrations make this the deepest ever  survey of the Galactic plane as seen from the northern hemisphere. Here we expect to discover many faint pulsars deep in the Galactic plane. This region is also where we have the greatest hope of discovering exotic systems such as double neutron stars and pulsars orbiting black holes, as these massive systems are expected to be close to their birth locations, given their relatively low ages \citep[see e.g.][]{Kramer2006}.

For all latitude regimes, the integration times were chosen such that the limiting flux densities of the HTRU-North and HTRU surveys \citep{Keith2010} were comparable. Table \ref{htrutab:strategy} shows integration times, data volumes and observational parameters for all parts of the HTRU-North survey. 

It should be noted that portions of the survey area of the HTRU-North are currently being searched in the Green Bank Northern Celestial Cap survey (GBNCC) with the Robert C. Byrd Green Bank Telescope \citep{Kaplan2012}, the Arecibo L-band Feed Array survey for pulsars (P-ALFA) with the Arecibo Observatory's William E. Gordon Radio Telescope \citep{Cordes2006} and several up-and-coming pulsar surveys with the Low Frequency Array (LOFAR). Despite the GBNCC and HTRU-North covering essentially the same survey region, the lower observing frequency of the GBNCC, 350 MHz, results in the surveys probing different subsets of the pulsar population.

\begin{figure}
\centering
\includegraphics[keepaspectratio=true,scale=0.4]{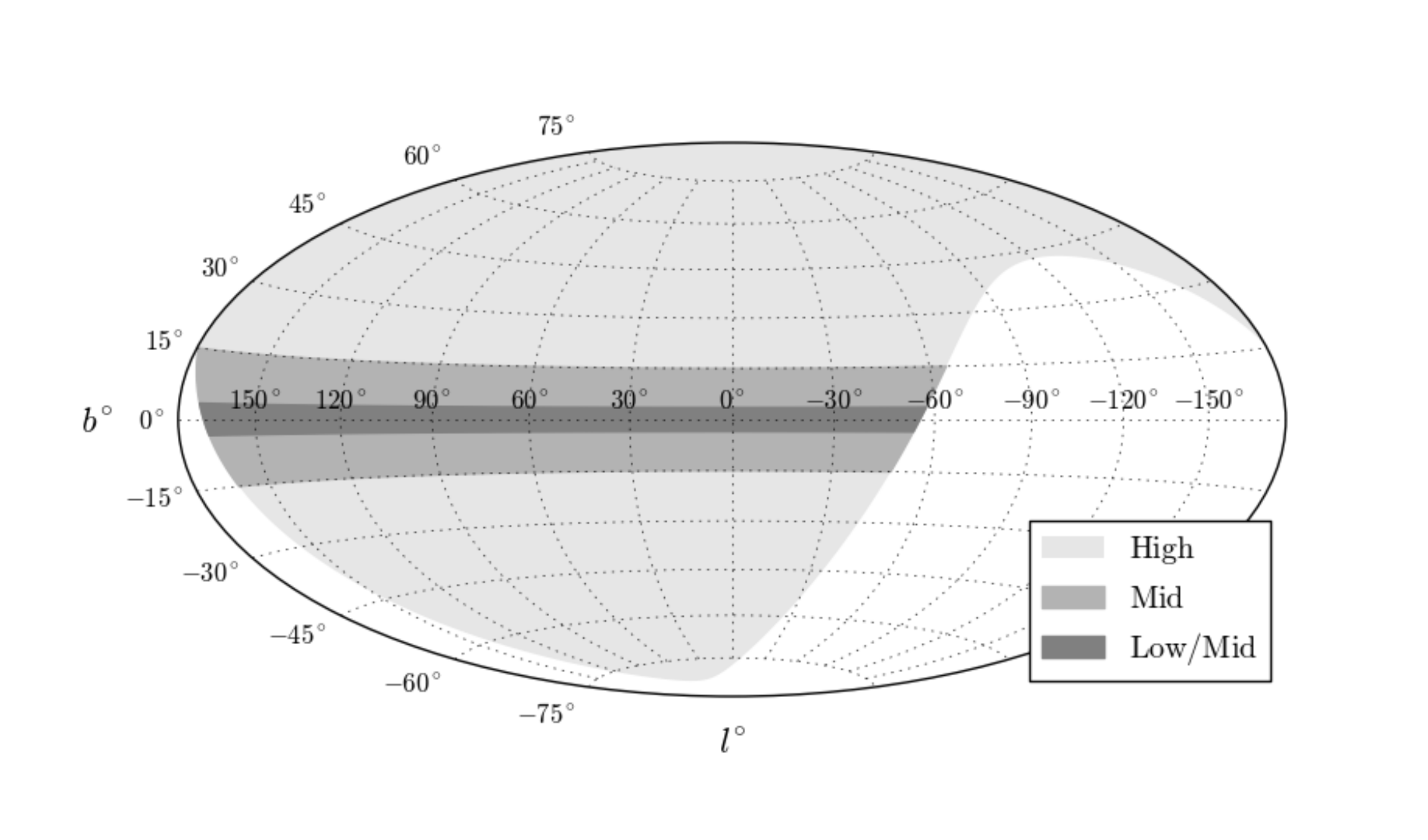}
\caption[]{A Hammer projection of the Galaxy showing the survey area for all three regions of the HTRU-North survey. Note that the mid-latitude portion of the survey also incorporates the low-latitude survey region. For information on the parameters of each latitude region, please consult Table \ref{htrutab:strategy}.\label{htrufig:strategy}}
\end{figure}

\begin{table}
\centering
\caption[Observing parameters of the HTRU-North pulsar survey]{Observational parameters for each latitude region of the HTRU-North survey. Both the bandwidth of the receiver ($\Delta\nu$) and the number of channels (N$_{\rm{chans}}$) are given post RFI removal. For each region the minimum detectable flux density (S$_{\rm{min}}$) for a pulsar with a 5\% duty-cycle was calculated using the method outlined in Section \ref{htru:sensitivity}. Here $G_{\rm{central}}$ and $G_{\rm{outer}}$ refer to the gains of the central in outer horns, respectively. Due to the physical location of the telescope, the survey has a minimum observable declination of $-20^{\circ}$.\label{htrutab:strategy}}
 \begin{tabular}[h]{lcccc}
 	\hline
    Survey & High & Mid & Low\\
    \hline
    Region & $|b| > 15^{\circ} $ & $|b|<15^{\circ}$ & $|b|<3.5^{\circ}$\\
    $t_{\rm{obs}}$ (s) & 90 & 180 & 1500 \\
    N$_{\rm{beams}}$ & 1,066,135 & 375,067 & 87,395\\
    $t_{\rm{samp}}$ ($\mu$s)& 54.61 & 54.61 & 54.61 \\
    $\Delta\nu$ (MHz) & 240 & 240 & 240 \\
    $\Delta\nu_{\rm{chan}}$ (kHz)& 585.9 & 585.9 & 585.9\\
    N$_{\rm{chans}}$ & 410 & 410 & 410\\
    	$G_{\rm{central}}$ (K Jy$^{-1}$) & 1.5 & 1.5 & 1.5\\
    	$G_{\rm{outer}}$ (K Jy$^{-1}$) & 1.3 & 1.3 & 1.3\\
    N$_{\rm{samples}}$ ($\times 10^{6}$) & 1.6 & 3.3 & 27.4\\
    	$S_{\rm{min}}$ (mJy) & 0.17 & 0.14 & 0.05\\
    Data/beam (GB) & 0.8 & 1.6 & 13.4\\
    Data (total) (TB) & 818.1 & 575.6 & 1117.8\\
    \hline
 \end{tabular}
\end{table}

\section{Instrumentation}
\label{htru:instrumentation}

All searching was performed using the 100-m Effelsberg radio telescope
of the Max-Planck-Institut f\"{u}r Radioastronomie. Below, we describe the receiver and backend systems used to
acquire observational data.

\subsection{The 21-cm Effelsberg multi-beam receiver}

The 21-cm Effelsberg multi-beam receiver consists of seven horns at
the prime focus of the Effelsberg telescope. The horns are arranged in
a hexagonal close-packed pattern around the central beam, with a beam
separation of 0.25$^{\circ}$. The central beam is circular with a
beamwidth (full-width half-maximum) of 0.16$^{\circ}$, while the outer beams have slight
ellipticity with a corresponding circular beamwidth of
0.166$^{\circ}$. Each of the seven horns has a bandwidth of 255 MHz centred on 1360 MHz and two polarisation channels, left- and right-hand circular for the central horn and orthogonal linear for the outer horns. Signals from the 14 channels are amplified in low-noise amplifiers, before undergoing down-conversion to an intermediate frequency of 150 MHz via heterodyning. After hardware RFI rejection we recover 250 MHz of useable band. This figure typically drops to $\sim$240 MHz after software RFI rejection (see Section \ref{htru:rfi_excision}).

During the course of an observation the parallactic angle of the beam
pattern on the sky changes with the telescope's azimuth-elevation position. To keep the
outer beams at constant Galactic latitude, the receiver box is rotated
to maintain a constant parallactic angle.

\begin{figure}
\centering
\includegraphics[width=240pt,height=220pt]{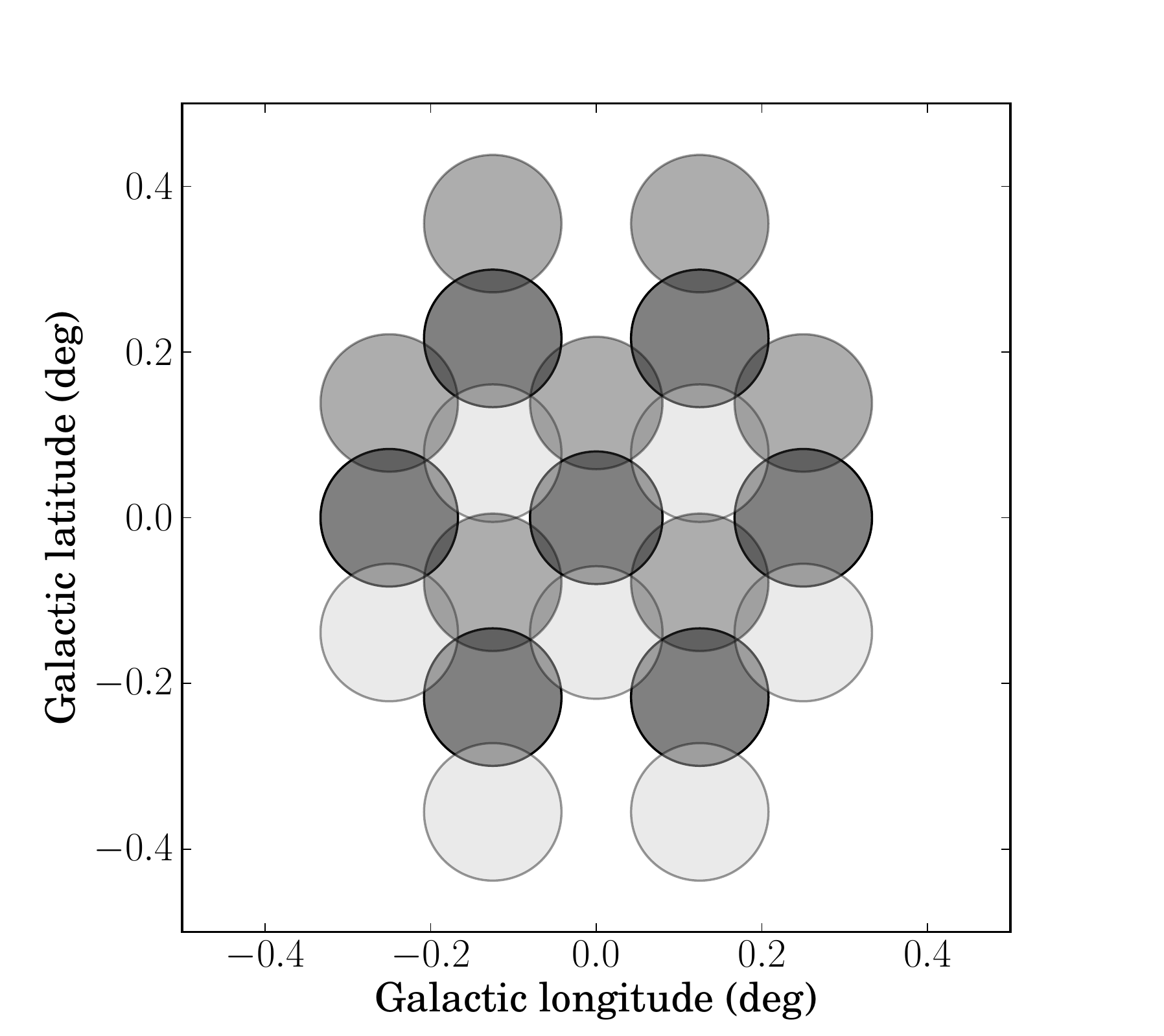}
\caption[The beam pattern of the 21-cm Effelsberg multi-beam receiver]{The beam pattern of the 21-cm Effelsberg multi-beam receiver on the sky. Shown are three pointings interleaved to make up a
  single tessellation unit for the survey. Circle diameters are given by the
  FWHM of each beam.}\label{htrufig:beampattern}
\end{figure}

The laboratory-measured receiver temperature of the central horn is 21 K, with the outer horns having temperatures of between 13 and 18 K.\footnote{The receiver temperatures for each horn, plus further information about the receiver system, can be found at http://www.mpifr-bonn.mpg.de/effelsberg.} The layout of the beam pattern on the sky and the tessellation unit for the
survey can be seen in Figure~\ref{htrufig:beampattern}.

\subsection{The PFFTS backend}

\begin{figure*}
\centering
\includegraphics[keepaspectratio=true,scale=0.5]{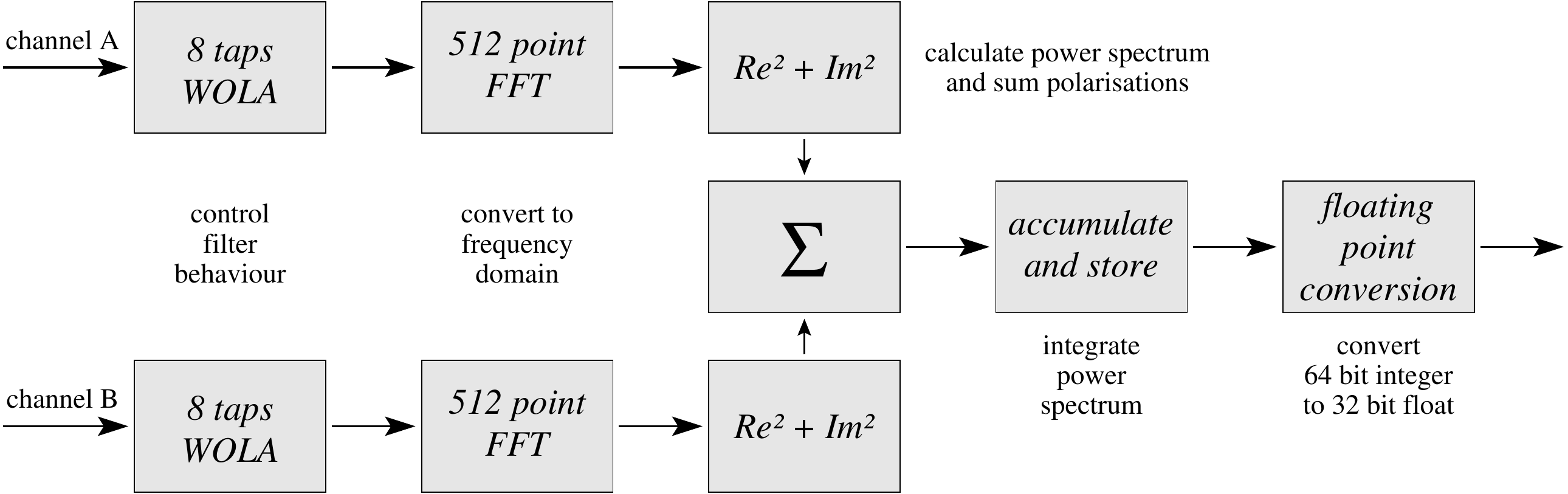}
\caption[PFFTS]{
   Block diagram of the FPGA signal processing pipeline.
   The digitised signals from both polarisation channels are individually transformed
   into the frequency domain by an 8-tap polyphase filter bank, implementing the weighted overlapp-add (WOLA) method, before conversion to a power density spectrum representation. The polarisations are then summed, integrated over an adjustable time-period and converted from 64-bit integers to 32-bit floating point numbers.
}\label{htrufig:pffts}
\end{figure*}

The Effelsberg Pulsar Fast-Fourier-Transform Spectrometer (PFFTS) backend was
specially developed to meet the requirements of searching for pulsars with a wide-band,
multi-beam receiver. The PFFTS is based on the AFFTS \citep{Klein2012} -- an FFT spectrometer
originally designed for spectral line observations. The backend combines seven identical 
electronic boards, each equipped with a high-speed analog-to-digital converter and
a high-performance field-programmable gate array (FPGA).
In ``pulsar search'' mode, the signal from each beam of the receiver is sampled by an analogue-to-digital converter (ADC) clocked at 600 MHz with 8-bit resolution. Following the ADC, the signal is processed by a 512-channel polyphase filter bank implemented on a Xilinx Virtex-4 FPGA. The output of the filter bank is further integrated by a factor of 32 and the two polarisations are quadrature summed.  Processing in the FPGA is done at 32-bit resolution in order to avoid numerical overflow, resulting in an aggregate data rate of $\sim$38 MB s$^{-1}$ per beam. Finally, the data are transmitted through the on-board gigabit ethernet controller to seven sever-class computers attached to the FPGA boards.

\section{Sensitivity}
\label{htru:sensitivity}
\subsection{Analytic sensitivity}

\begin{figure}
\centering
\includegraphics[width=250pt,height=250pt]{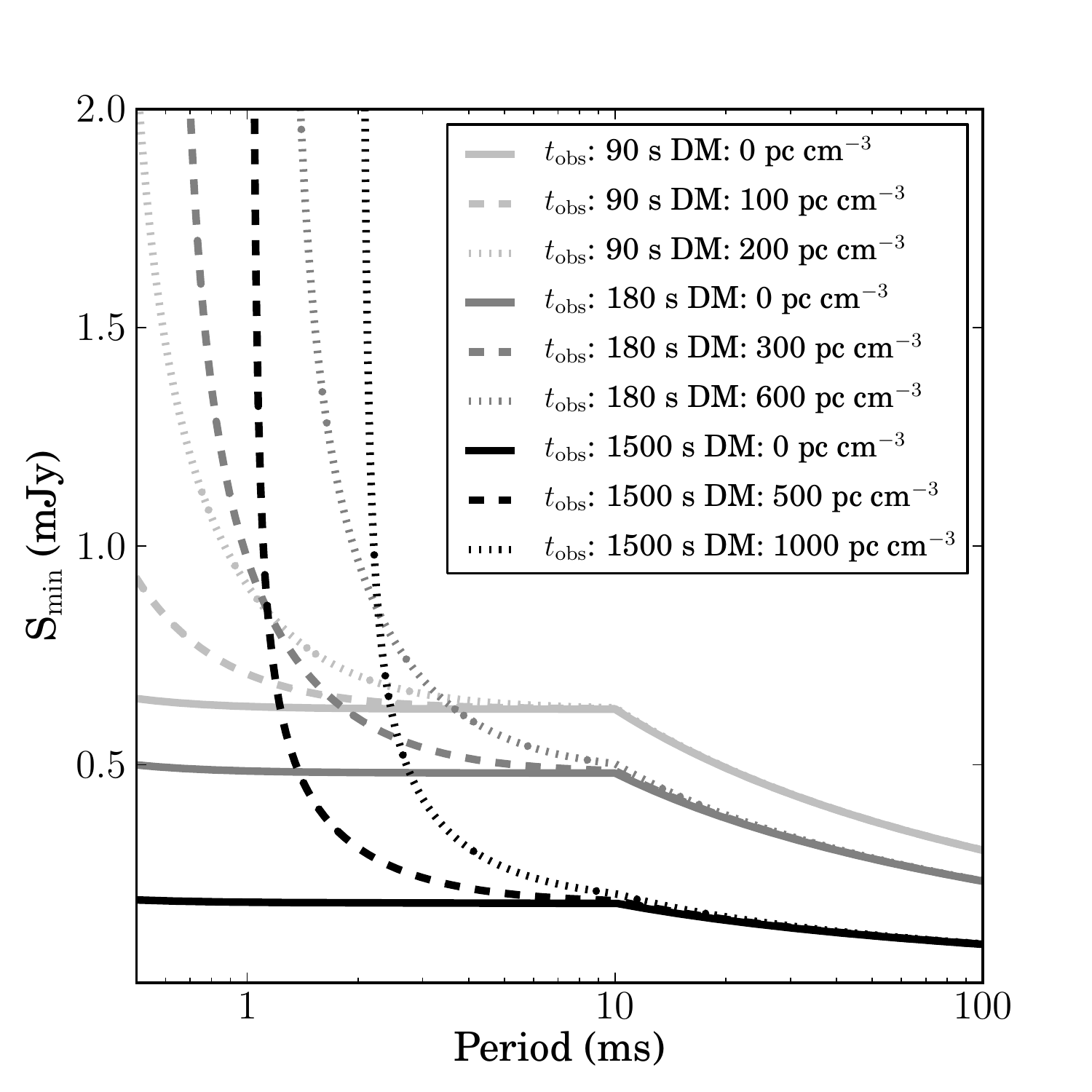}
\caption[Estimated sensitivities for each region of the HTRU-North pulsar survey]{Theoretical minimum detectable flux densities
  ($S_{\rm{min}}$) vs$.$ spin period for high-latitude, mid-latitude
  and low-latitude regions of the survey. For each region, the lines
  from left to right represent increasing dispersion measures. As the value of  $S_{\rm{min}}$ is dependent on the duty-cycle ($\delta = W_{\rm{eff}}/P$) of the pulsar, we use the relation $\delta=P^{-1/2}$ \citep{Kramer1998} to describe a more realistic duty-cycle distribution. As noted by \citet{Kramer1998}, this relation does not hold for short-period pulsars. To account for this, a maximum duty-cycle of $1/3$ is imposed, resulting in the breakpoint seen at $\sim10$ ms. For each integration length the sky temperature, $T_{\rm{sky}}$, is given by the average sky temperature for the corresponding latitude range; 11 K for low latitudes, 8 K for mid latitudes and 5 K for high latitudes.  A minimum detectable S/N ratio of 8 is
  imposed.}\label{htrufig:sensitivity}
\end{figure}

To estimate the minimum pulsed flux density, $S_{\rm{min}}$, observable by our survey, we use the modified radiometer
equation \citep[see e.g.][]{handbook},
\begin{equation} \label{htrueq:sensitivity}
S_{\rm{min}} = \beta\frac{\rm{S/N}_{\rm{min}} \mathit{T}_{\rm{sys}}}{G \sqrt{n_p
    t_{\rm{obs}} \Delta f }} \left (
\frac{W_{\rm{eff}}}{P-W_{\rm{eff}}} \right 
)^{\frac{1}{2}} ,
\end{equation}
where the constant factor $\beta$ denotes signal degradation caused by
digitisation, which for 8-bit sampling is $\sim0.01\%$, giving
$\beta \approx 1$ \citep{Kouwenhoven2001}. The system temperature, $T_{\rm{sys}}$, is the sum of the receiver temperature, $T_{\rm{rec}}$, and the sky temperature $T_{\rm{sky}}$. From flux density calibration measurements we find
$T_{\rm{rec}} = 21$ K for the central horn. The antenna gain, $G$, is 1.5 K Jy$^{-1}$ at 1.36 GHz. 
Other parameters in this expression are the total integration time, $t_{\rm{obs}}$; the effective bandwidth of the receiver, $\Delta f$;  the number of polarisations summed, $n_p$, which for this survey is always two; the pulsar period, $P$; the effective pulse width, $W_{\rm{eff}}$;  and the minimum S/N ratio with which we can confidently make a detection, $\rm{S/N}_{\rm{min}}$. Based on false alarm statistics \citep[see e.g.][]{handbook}, $\rm{S/N}_{\rm{min}} = 8$. Due to intra-channel dispersive smearing the effective pulse width increases with dispersion measure\footnote{Here, the dispersion measure is defined as the integrated column density of free electrons along the line of sight between observer and pulsar.} (DM) as
\begin{equation}
W_{\rm{eff}} = \sqrt{W_{\rm{int}}^2 + \left(k_{\rm{DM}}\frac{\Delta f_{\rm{chan}}}{f^3} {\rm{DM}}\right)^2 + \mathit{t}_{\rm{samp}}^2}, 
\end{equation}
where $W_{\rm{int}}$ is the intrinsic pulse width, $t_{\rm{samp}}$ is the sampling interval of the observation, $f$ is the observing frequency, $\Delta f_{\rm{chan}}$ is the bandwidth of a single frequency channel and $k_{\rm{DM}} = 8.3\times10^3$ s. While scatter broadening will likely be the limiting factor on the DM depth to which we can detect short-period pulsars, the large uncertainty on its relationship with DM \citep{Bhat2004} means that we disregard it when calculating expected sensitivities. For this reason, the sensitivity calculations here represent a best-case scenario. Figure \ref{htrufig:sensitivity} shows sensitivity curves for each region of the survey for a selection of DMs. 

\subsection{Pulsar redetections}
\label{htru:redetections}

Thus far, observations have been concentrated on the mid-latitude portion of the HTRU-North survey.
Processing of the first 13\% of these observations has led to the redetection of 93 known pulsars. To obtain an empirical confirmation of our survey sensitivity, observed S/N ratios were compared to
S/N ratios predicted using published flux densities taken from the
ATNF pulsar catalogue\footnote{http://www.atnf.csiro.au/research/pulsar/psrcat/ \citep{Manchester2005}}.

Using the sky temperature model of \citet{Haslam1982}, scaled with a
spectral index of $-2.6$ \citep{Lawson1987}, and published pulsar positions taken from
the ATNF pulsar catalogue, we calculated the expected S/N ratio for all redetections
through rearranging equation \ref{htrueq:sensitivity}. As the redetections
did not lie in the centre of their discovery beam, S/N values were
multiplied by a Gaussian offset factor, $q = e^{-(\theta/\phi)^2\ln{0.5}}$, to correct for off-axis gain decreases. Here
$\theta$ is the pointing offset in degrees and $\phi$ is the beam
half-width at half maximum.

\begin{figure}
\centering
\includegraphics[keepaspectratio=true,scale=0.45]{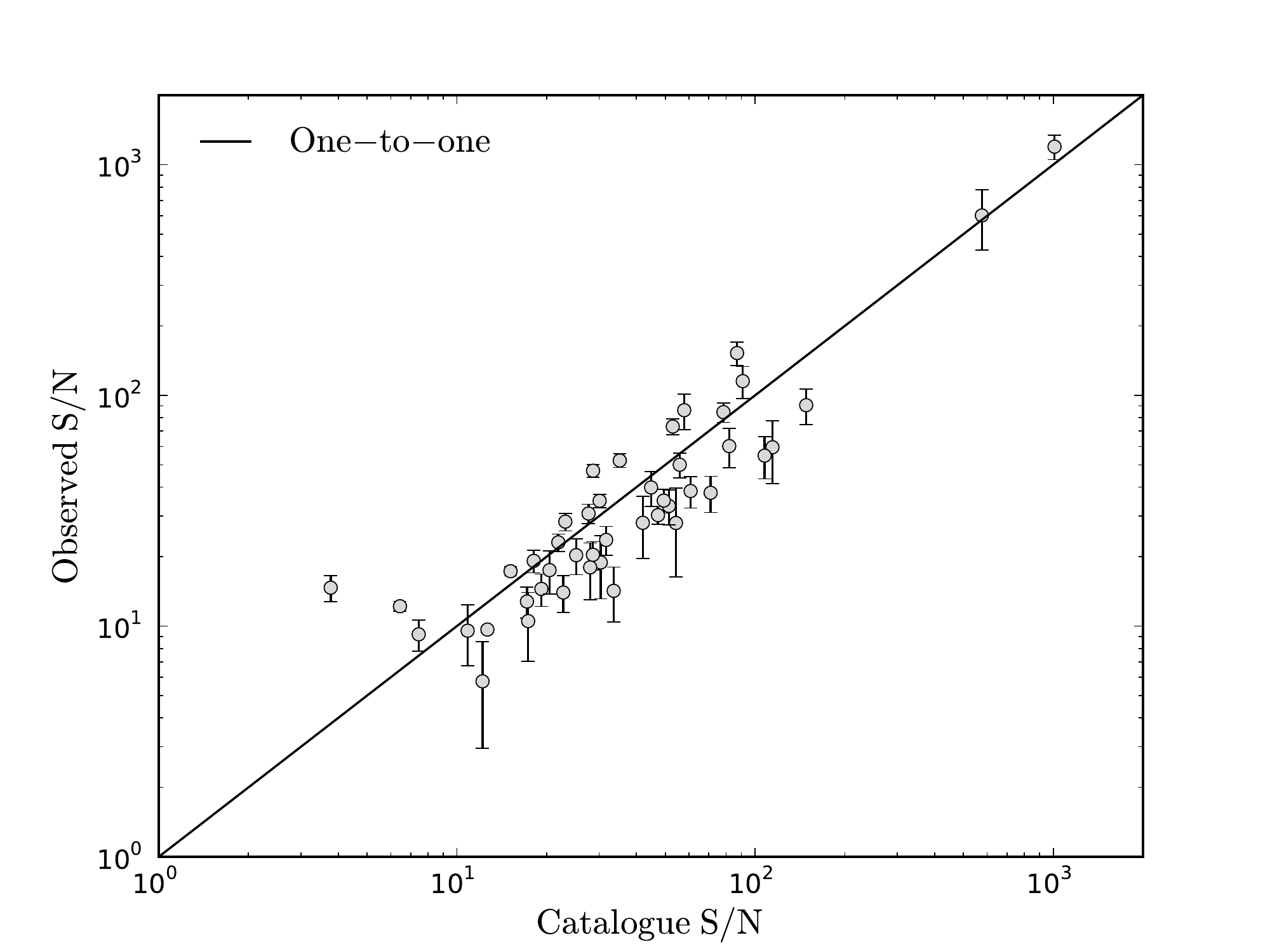}
\caption[Analysis of pulsar redetections from the HTRU-North pulsar survey]{Observed S/N vs$.$ expected S/N for redetected HTRU-North pulsars. Estimated S/N values are calculated using
  published flux densities taken from the ATNF pulsar catalogue, with
  a scaling factor for the offset of the pulsar from the bore-sight of
  the beam.}\label{htrufig:redetections}
\end{figure}

To make the comparison more robust, only redetections that were within
one beamwidth of the observed position were used. Redetection observations were also cleaned of RFI prior to S/N measurement (see Section \ref{htru:rfi_excision}). Figure \ref{htrufig:redetections} shows the observed S/N versus the expected S/N for the remaining sample of redetections.

As an independent test for sensitivity losses in the backend, timing
observations from the Lovell telescope were also used to obtain S/N
ratio measurements for the newly discovered pulsars from this survey. In all
cases the S/N ratio measurement from the Effelsberg discovery observation
agreed with the distribution of S/N ratio measurements obtained from the Lovell
telescope timing data to within 1$\sigma$.

To verify that no pulsar had been missed by the survey, all pointings
for which a known pulsar was within one beamwidth were examined. S/N
ratios for these pulsars were estimated using the method outlined
above. Of the pulsars with estimated S/N ratios above 8, five were
undetected in the initial processing of the data. Folding the data for
these pulsars with published ephemerides led to detections for two of
the pulsars with S/N ratios below our detection limit. The
three remaining undetected pulsars all have periods in
excess of $2.5$ s. Long period systems such as these, are often difficult
to detect in short observations, as wider Fourier bins and large
low-frequency components in the data act to suppress the pulsar signal. It
should be noted that although these three pulsars were undetected in
periodicity searches, one was detected through single-pulse analysis.
These results confirm that the observing system is performing as expected,
with no loss of sensitivity.

\section{Simulations}
\label{htru:simulations}
To estimate the expected detection rate of the HTRU-North survey, Monte-Carlo
simulations of the Galactic MSP and normal pulsar populations were
performed using the model outlined in \citet{Lorimer2006}, with the
\textsc{PSRPOP}\footnote{http://psrpop.phys.wvu.edu/} software.  To
simulate the normal pulsar population, input model parameters were
chosen as follows:

\begin{itemize}
\item Empirical period distribution taken from the probability density function of the known
population.
\item A log-normal pseudoluminosity distribution, defined at 1.4 GHz, with mean and standard deviation in
  log space of $-1.1$ and $0.9$, respectively \citep{FaucherGiguere2006}.
\item An exponential distribution for the height above the Galactic
  plane, with a scale height of $330$ pc \citep{Lorimer2006}.
\item A radial distribution as described in \citet{Lorimer2006}.
\item A 6\% duty cycle with dither given in \citet{Lorimer2006}.
\item The NE2001 Galactic free electron density model \citep{Cordes2002}.
\end{itemize}

The number of pulsars to be simulated was chosen such that the discovery rates of
simulated versions of the Parkes Multibeam Pulsar Survey
\citep{Manchester2001}, the Swinburne Intermediate Latitude Pulsar
Survey \citep{Edwards2001} and its extension \citep{Jacoby2009} and
the Parkes High Latitude Survey \citep{Burgay2006a}, matched those of
their real counterparts. 

To estimate the number of MSP detections expected from the survey, we used model `A' of \citet{Lorimer2012}. This model updates the results of \citet{Lorimer2006} by taking into account Galactic MSPs found through recent high time- and frequency-resolution pulsar searches. Table \ref{htrutab:simulations} shows the results of the simulations and the expected discovery count for each region of the HTRU-North survey. The simulations suggest that the HTRU-North survey will detect $\sim1657$ normal pulsars and $\sim153$ MSPs, after taking into account pulsars co-detected in both the mid- and low-latitude portions of the survey. However, recent work on determination of the MSP luminosity distribution using pulsar detections from the HTRU survey (Levin et. al, in prep.) has suggested that larger MSP yields may be expected.

\begin{table}
\centering
\caption[Simulated yields of the HTRU-North pulsar survey]{Simulated results for the total number of pulsars detected in each latitude region of the HTRU-North survey. To estimate the number of new discoveries from each region, the number of known pulsars expected to be detected in that region was subtracted from the simulated detection count.} \label{htrutab:simulations}
 \begin{tabular}[h]{lllll}
 \hline	
 	\multicolumn{1}{c}{} & \multicolumn{2}{c}{Detections} & \multicolumn{2}{c}{Discoveries}\\
  \multicolumn{1}{c}{Region} & \multicolumn{1}{c}{Non-MSPs} &
\multicolumn{1}{c}{MSPs} & \multicolumn{1}{c}{Non-MSPs} &
\multicolumn{1}{c}{MSPs}\\
  \hline
  High-lat & 145 & 28 & 29 & 7\\
  Mid-lat & 784 & 66 & 142 & 41\\
  Low-lat & 1123 & 81 & 642 & 64\\
  \hline
 \end{tabular}
\end{table}

\section{Data analysis}
\label{htru:processing}

Processing of data collected for the  HTRU-North survey is currently performed both
on-site at the Effelsberg observatory and at the Max-Planck-Institut
f\"{u}r Radioastronomie in Bonn. Data undergo pre-processing and RFI
treatment before being processed twice, once in a `quick-look' pipeline that operates on reduced time- and frequency-resolution data and is sensitive to the majority of isolated pulsars in the
data, and once in a full pipeline that incorporates searches for
pulsars in compact binary systems. Below we describe all stages in the
processing and archiving of data from the HTRU-North survey. It should be noted
that the data analysis procedure reported here is only valid for
mid- and high-latitude pointings from the survey. The analysis
procedure for the low-latitude pointings will be presented elsewhere.

\subsection{Pre-processing}

Initially, data written in 32-bit format by the PFFTS backend are
down-converted to 8-bit format for storage, transportation efficiency and
software compatibility. During the conversion, the data in each
frequency channel are clipped at the 3-$\sigma$ level\footnote{i.e. all data points with values $>3\sigma$ (three standard deviations above the channel mean) are reduced in power such that they have a value equal to the 3-$\sigma$ level of the original data.}, allowing the data to be mapped to 8 bits with minimal loss in dynamic range. A
by-product of this process is that the bandpass shape is removed from the data, as noisy channels are down-weighted
and quiet channels are up-weighted with respect to one another. For
purposes of RFI mitigation and completeness in the data archiving
system, the original 32-bit data bandpass shape is stored.

\subsection{RFI excision}
\label{htru:rfi_excision}
Before the data are searched for pulsars, they are treated with several
RFI excision methods to remove spurious signals of man-made origin.
In the first stage of RFI removal, frequency channels with average power
levels 3$\sigma$ or more above the normalised mean across the original
32-bit bandpass are replaced with zeros. This reduces our sensitivity to
weak, persistent, narrow-band RFI.
The multibeam nature of the receiver allows for the application of a
spatial filtering system to mitigate impulsive RFI in the
data. Assuming all RFI that enters the multi-beam receiver is
temporally coherent, we may apply a simple thresholding scheme to each
data point to identify interference which appears in multiple
beams.

The PFFTS is an adapted version of the AFFTS backend \citep{Klein2012} used for HI observations at Effelsberg. As HI observations with the multibeam receiver do not require the high time resolution of pulsar observations, the AFFTS was not designed to have accurate synchronisation between individual beam servers. This can result in lags of up to a few milliseconds between the start of recording between different beams of the same pointing. Therefore, to perform multi-beam impulsive RFI excision, data must be cross-correlated to determine the absolute time offset. This process
inherently relies on the presence of a multi-beam signal in the data which will
produce a strong feature in the cross-correlation. In the cases where no such
signal exists the impulsive masking section of the RFI excision is
bypassed.

After an absolute reference has been determined for each pointing,
each data point is compared across the seven beams. If the data point
has a significance of $1.5\sigma$ in four or more beams, it is
replaced with Gaussian noise indistinguishable from the surrounding
data and is logged for further analysis. Assuming that each
channel is composed of Gaussian noise, the chance probability of removing
a single `good' data point is 0.00013\%. The value of $1.5\sigma$, as with all threshold values used in the RFI mitigation procedure, was chosen based on empirical tests of the mitigation procedure.

\begin{figure}
\includegraphics[scale=0.55]{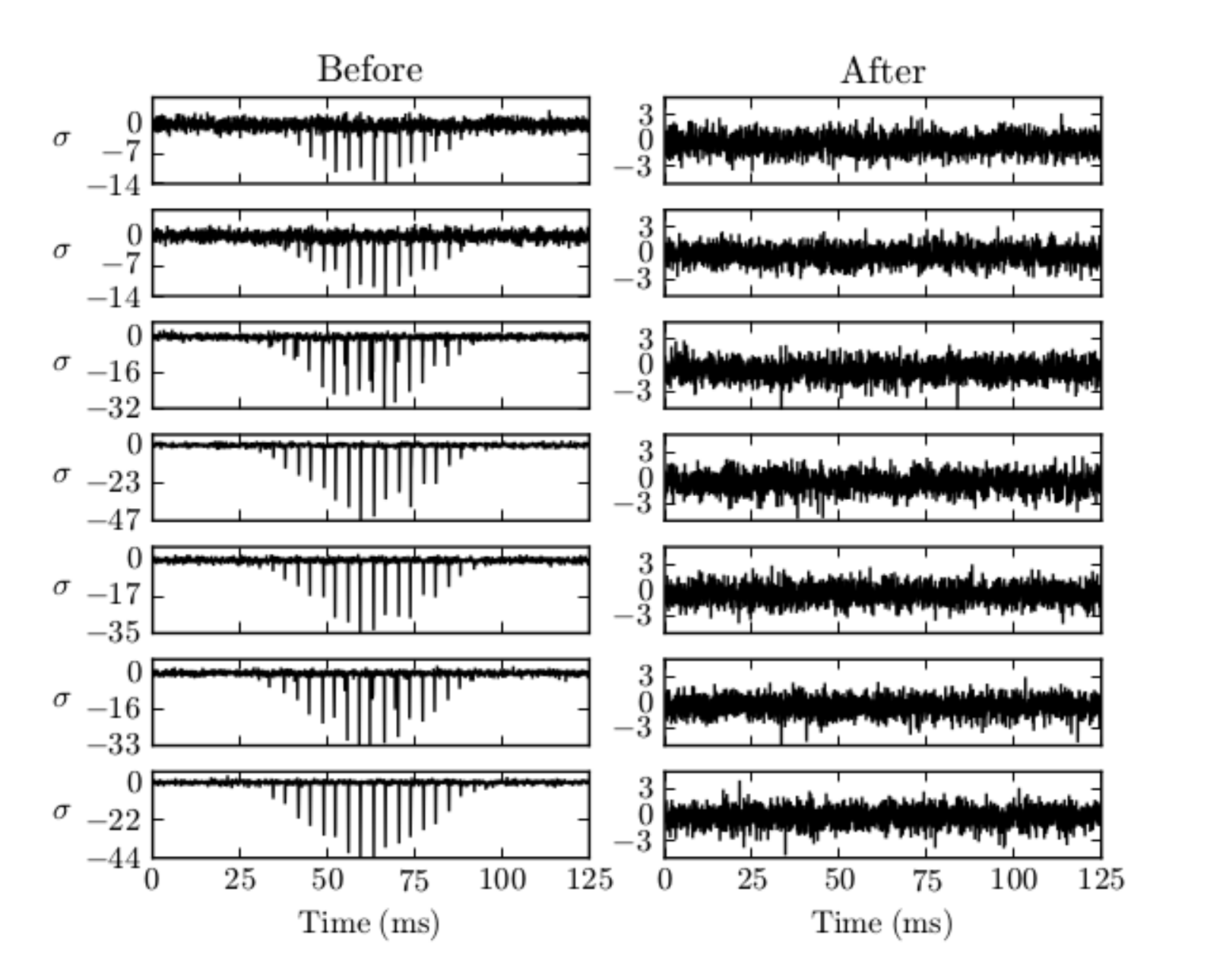}
\caption{A the zero-DM time series for each beam of the multibeam receiver prior to (\textit{left-hand column}) and post (\textit{right-hand column}) RFI excision. The y-axis for each panel shows the number of standard deviation from the mean of the data. Here, the RFI appears as a decrease in power due to a combination of the low-noise amplifier response for very bright, narrowband signals and the clipping of each channel at the 3-$\sigma$ level during conversion of the data from 32-bit floats to 8-bit integers.\label{htru:rfi_plot}}
\end{figure}

Once all beams have been compared, histograms of the RFI-flagged data points are created, both by time sample and frequency channel. By examining the number of RFI-affected data points in each frequency channel, we
can isolate channels which have persistent impulsive noise. If the percentage of RFI-affected data points in a given channel is greater than 0.2\%, the data in that channel are replaced by
zeros. Similarly if more than 10\% of the channels in a
given time sample are RFI-affected, then all channels in the time sample
are replaced by Gaussian noise, unless they have been previously
replaced by zeros. An example of the zero-DM time series for each beam of the receiver before and after impulsive RFI excision can be seen in Figure \ref{htru:rfi_plot}. 

To identify periodic signals in the pointing, the `clean' data are
collapsed along their frequency axis, with the resultant `zero-DM' time series
analysed in the Fourier domain (see Section \ref{htru:periodsearch}). Fourier frequencies which appear in
four or more beams with a 2-$\sigma$ significance or greater are
written to a `zaplist' file that is used during periodicity
searching and candidate sorting in the processing pipelines. 

Typically the RFI excision process removes $\sim4\%$ of the raw data. The majority of this is due to removal of the band edges and supression of noisy frequency channels. After removal of these channels we recover and effective bandwidth of $\sim$240 MHz.   

\subsection{Processing pipeline}

Here we cover the main stages of the full processing pipeline used to analyse
HTRU-North data. The pipeline is built around the \textsc{presto} data
analysis package \citep{Ransom2001}. The quick-look pipeline is described in Section \ref{htru:quicklook}. 

\subsubsection{De-dispersion}
\label{htru:dedispersion}
Broadband electromagnetic signals propagating through the interstellar medium are subject to group-velocity dispersion. This results in a frequency-dependent time delay in the signal, with components at higher frequencies arriving at the observer before those at lower frequencies. As the degree to which the signal from an
unknown pulsar is dispersed (its dispersion measure, DM), is not known
\textit{a priori}, we search 3240 trial DMs in the range 0-978 pc
cm$^{-3}$. Such a large number of trials allows for retention of
the data's highest possible time resolution at all DMs, as the minimum time delay between adjacent trials is limited only by the sampling rate and time delay between the top and bottom of a single frequency channel. It should be noted that to account for high-DM pulsars and transients the quick-look pipeline searches up to a maximum DM of 3000 pc cm$^{-3}$ (see Section \ref{htru:quicklook}).
At this stage in the analysis, the data are barycentred to remove the effects of
the Earth's rotation and motion in the Solar System.

\subsubsection{Periodicity searching}\label{htru:periodsearch}

Each of the 3240 time series created in the de-dispersion stage of the pipeline
must be searched for periodic signals from isolated pulsars and pulsars in
binary systems. To this end, the time series are discrete-Fourier-transformed
to create a power spectrum for each DM trial. Often the power spectra
contain strong low-frequency noise from long-period RFI or gain
fluctuations in the receiver. To mitigate this, the power spectra are
de-reddened through subtraction of an interpolated red-noise curve \citep{Israel1996}.
At this stage, Fourier frequencies which have been found to contain RFI through the excision process are suppressed in the spectra.

To reconcentrate power distributed through harmonics in the Fourier domain, the
process of incoherent harmonic summing is used. 
Here, the spectrum is summed with a stretched copy of itself such that all second harmonics are added to their corresponding
fundamentals. This process is repeated four times such that all power distributed
in even harmonics up to the 16$^{\rm{th}}$ harmonic may be added to the fundamental
\citep[see e.g.][]{handbook}. To identify non-accelerated signals in the data, the spectra from each
stage of the harmonic summing are searched for significant peaks.

In the case of pulsars in binary systems, the Doppler effect causes the
apparent spin frequency of the pulsar to drift over the course of an observation, spreading the pulsar's
power in the Fourier domain. To reconstruct power
smeared across multiple Fourier bins, we employ the `correlation technique' of
matched filtering as outlined in \citet{Ransom2002}.
The number of Fourier bins drifted by the signal, $N_{\rm{drift}}$,
and the binary acceleration of the pulsar, $a_{0}$, are related by
$a_{0} = N_{\rm{drift}}Pc/t^{2}_{\rm{obs}}$, where $P$ is the spin-period of the pulsar, $c$ is the speed of light and $t_{\rm obs}$ is the observation length. To achieve sensitivity to
accelerations of $|250|$ m s$^{-2}$ for a 1-ms pulsar, we search
values of $N_{\rm{drift}}$ up to 27 for medium-latitude pointings and
7 for high-latitude pointings. Although the current value of $|250|$ m s$^{-2}$ is high enough to enable the survey to detect the most relativistic binary systems currently known, our assumption of a constant line-of-sight acceleration reduces our sensitivity to orbits that have $P_{\rm{b}}\lesssim10t_{\rm{obs}}$ \citep{Ransom2003}.

\subsubsection{Candidate sifting}

To reduce the large quantity of candidates that periodicity searching produces, we apply a selection of thresholds and excision criteria. Initially, signals with the same period at different DMs are combined into a single candidate. At this stage any duplicate candidates are removed. As the relationship between S/N degradation and $\Delta$DM (where $\Delta$DM is the the offset from a pulsar intrinsic DM) for pulsars is well understood, we can excise candidates based on their DM characteristics. Candidates that are strongest at a DM lower than 2 pc cm$^{-3}$ or do not show up at two or more consecutive DM trials are removed. Candidates that are lower significance harmonics of other candidates are also removed. Finally all candidates are sorted by significance to be passed to the folding algorithm. 

It should be noted that all candidate sifting is done on a single-beam basis and no information from the other beams of the same pointing is used in determining the most pulsar-like candidates. Although it adds an extra layer of pipeline complexity, it is expected that future re-processings of the survey will benefit from implementing spatial filtering techniques to reduce the number of spurious candidates from each pointing.

\subsubsection{Folding and optimisation}

To determine if a candidate is truly a pulsar, the data are phase-folded and de-dispersed at the period and DM found through candidate
sifting. After the data are folded, both the period and the DM of the
candidate can be optimised through searching a small range of values
around the discovery values. The optimisation is tailored such that
for faster-period candidates, smaller ranges in DM and period are
searched. To reduce sensitivity to RFI, long-period candidates do not
undergo DM optimisation.

All candidates with greater than 8-$\sigma$ significance are folded. In beams where there are less than 50 candidates above 8-$\sigma$ significance, candidates with greater than 6-$\sigma$ significance are also folded such that there is a minimum of 50 folds. 
Limiting the number of folds in this way reduces the probability of missing pulsars in observations strongly affected by RFI, as all candidates with a significance above our detection threshold are folded. 

\subsubsection{Candidate viewing and ranking}
\label{htru:candranking}
To deal with the $>$80 million candidates the survey will produce, a
suite of interactive plotting software coupled with a
\textsc{MySQL}\footnote{www.mysql.com} database has been developed. For each
folded candidate, the database stores all the relevant statistics of
that candidate.  Through use of the viewing software, users may query
the database to select candidates which satisfy certain criteria,
before viewing those candidates in the parameter space of their
choice. User rankings of each candidate are stored in the database,
with highly ranked candidates marked for re-observation.

For someone with experience in candidate selection, taking on average
two seconds to view each candidate, it would take five years without
pause to view all candidates produced by the survey. To reduce the
volume of candidates that must be inspected, we implement both an
artificial neural network (ANN) and an automatic ranking algorithm in
post-processing.

The \textsc{peace} (Pulsar Evaluation Algorithm for Candidate
Extraction) software package (Lee et al. 2013, submitted) is used to generate automatic rankings for each candidate. Here, the software weights and combines a selection of scores, determined through analysis of the folded data, to generate an overall `likeliness-of-pulsar' measure for each candidate. As the \textsc{peace} software is designed to detect pulsars that display expected properties, it is subject to selection bias against atypical systems.

ANNs are a class of computational techniques which attempt to emulate
the decision making behaviour of a human mind. ANN have been
successfully applied to candidate selection
\citep[e.g.][]{Eatough2010,Bates2012} and have been shown to reduce
the number of candidates required to be looked at by several orders of
magnitude.  To train the ANN, it is provided with a vector of
`scores', in this case generated by the \textsc{peace} software, for each candidate from a selection of both real and simulated pulsar signals and RFI. The use of ANNs must also be treated with care, as their
sensitivity to pulsars which do not exhibit typical behaviour
(e.g. pulsars which are intermittent, in binary systems or highly
scintillating) is dependent on the composition of the data set
used in training.
Although both ANNs and \textsc{peace} are effective in determining whether a
candidate is a pulsar or not, direct visual inspection of the
candidates is still the primary method of pulsar identification. The
rankings generated through visual inspection act as an absolute
reference for all automatic ranking systems.

\subsection{Quick-look pipeline}\label{htru:quicklook}

The aim of the quick-look pipeline is to perform a reduced version of the full pipeline that is capable of processing all acquired data between observing sessions. By keeping
up-to-date with the observed data, we are able to monitor the
performance of the receiver and backend systems, as well as
maintain up-to-date knowledge of the RFI environment at the Effelsberg
telescope.

Data passed to the quick-look pipeline are initially downsampled by factors of
four and two in time and frequency, respectively. Although downsampling reduces
our sensitivity to short-period/high-DM pulsars, it increases the
throughput of the pipeline by a factor of eight, allowing processing to be
performed with limited resources at $1.6 \times$ real-time. The data are
de-dispersed to 406 trial dispersion measures in the range 0-3000 pc cm$^{-3}$. To
perform multiple de-dispersions efficiently, we employ the method outlined in
\citet{Keith2010}, with the caveat that we must first reduce the data to 7-bit
resolution to avoid integer overflow in the output data.

For each trial DM, two searches are performed: a Fourier-domain search for
periodic, unaccelerated signals and a time-domain search for isolated pulses.
The Fourier-domain search closely follows the methodology outlined for the full
pipeline with the exception that no acceleration searching is performed.

\subsubsection{Transient searching}

To search for isolated pulses in the time domain, we follow a
similar methodology to that outlined in \citet{Burke-Spolaor2010}. Here, we use
matched filtering to identify significant impulsive signals of varying widths.
Signals with significance greater than 4$\sigma$ are collated and compared
across all DMs to determine whether a candidate obeys the cold plasma dispersion relation and to determine
that candidate's optimal dispersion measure. As the data have already undergone spatial coincidence filtering during pre-processing, no event matching is required across beams.

Candidate detections from the transient search are viewed on a pointing-by-pointing
basis, with interesting signals being followed-up using software that provides tools for interactive manipulation and viewing of the filterbank data. Follow-up in this manner is vital in determining if a signal is of astrophysical origin or is simply RFI. Although the transient search has detected many known pulsars, no previously unknown transients have so far been discovered. 

The remaining steps of the quick-look pipeline follow the same process as the
full pipeline, with the exception that candidates are stored independently of the
\textsc{MySQL} database.

\section{New pulsar discoveries}
\label{htru:discoveries}

Here we present the initial pulsar discoveries of the HTRU-North
survey. So far the survey has discovered 15 pulsars including one MSP. All discoveries originate from the processing of the first 13\%
of the mid-latitude region of the survey. 

\begin{figure*}
\centering
\includegraphics[keepaspectratio=true,scale=0.55]{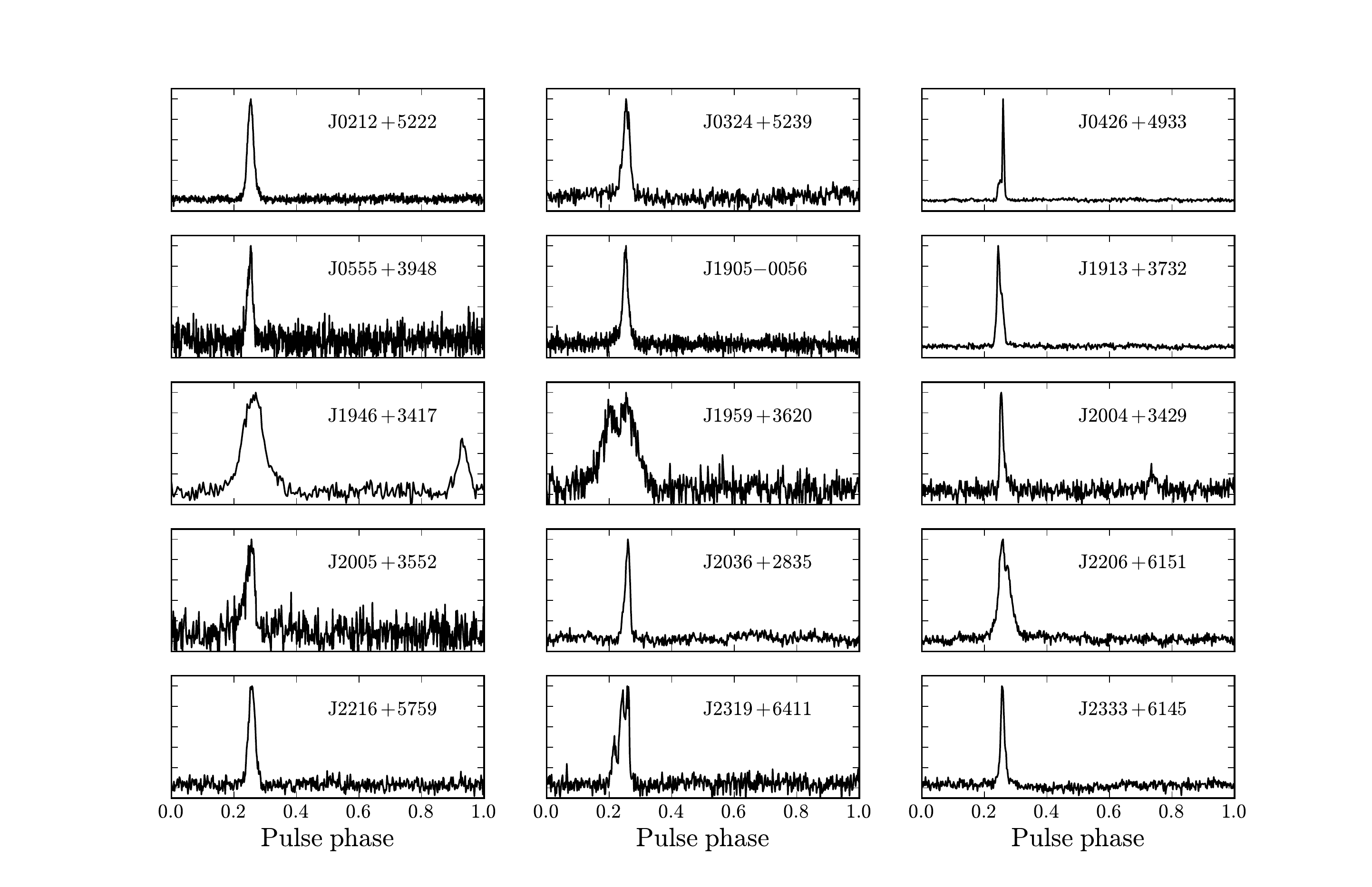}
\caption[Pulse profiles of the 15 newly discovered pulsars from the HTRU-North pulsar survey.]{Integrated pulse profiles for the 15 newly discovered pulsars of the HTRU-North survey. The y-axes show the flux density of each pulsar on arbitrary scales. For the mean flux density of each pulsar, please refer to Table~\ref{htrutab:derived}.\label{htrufig:profs}}
\end{figure*}

Upon discovery, each new pulsar is timed by the Lovell radio telescope
at Jodrell Bank observatory and the Effelsberg radio telescope. Pulse times-of-arrival (TOAs) are
analysed with the \textsc{tempo2} software package \citep{Hobbs2006}
to create phase-connected timing solutions for each pulsar. Tables
\ref{htrutab:eph} and \ref{htrutab:derived} show the new pulsar discoveries with current
timing solutions and derived properties. In the case of PSR J0555$+$3948 the short timing baseline precludes the accurate determination of both position and flux density. In this case, the position error is assumed to be equal to the half-width half-maximum of a single beam and the flux density measurement assumes the discovery position to be the true position. As period derivative ($\dot{P}$) and position are covariant over short timing baselines, the $\dot{P}$ measurement for all pulsars with data spans smaller than one year should be treated with caution. Figure \ref{htrufig:profs} shows integrated pulse profiles from coherently dedispersed observations with the Lovell telescope. All timing observations are conducted in the 21-cm band with centre frequencies of 1.36 and 1.53 GHz, and bandwidths of 400 and 200 MHz for the Lovell and Effelsberg radio telescopes, respectively. 

Observations with the Lovell are performed $\sim2$-$3$ times per week until a preliminary timing solution for the pulsar can be determined. The cadence of observations is then reduced to $\sim1$ observation every three weeks. After the pulsar's position has been improved through continued timing with the Lovell, higher precision timing observations with Effelsberg begin. Effelsberg observations for each pulsar occur on a monthly basis.

\begin{table*}
\centering
\caption[Timing solutions for 15 pulsar discovered by the HTRU-North pulsar survey.]{Timing solutions for the first discoveries of the HTRU-North survey. Numbers in parentheses represent twice the formal 1-$\sigma$
  uncertainties in the trailing digit as determined by
  \textsc{tempo2}. Here, $\dot{P}$ is the first derivative of the pulsar's spin-period. All parameters are measured w.r.t$.$ reference epoch MJD 56100, unless otherwise stated. These parameters were determined with \textsc{tempo2}, which uses the
  International Celestial Reference System and Barycentric Coordinate Time. Refer to \citet{Hobbs2006} for information on modifying this timing model for observing systems that use \textsc{tempo} format
  parameters.\label{htrutab:eph}}
\begin{tabular}{lllllllllll}
\hline
PSR & R.A.          & Decl.             & $P$ & $\dot{P}$ & DM & $N_{\rm{TOA}}$ & Data span & Residual \\
    & (h:m:s) & ($^{\circ}:':''$) & (ms)   & ($10^{-15}$)  &   (pc cm$^{-3}$)    &    &  (MJD)  & ($\mu$s)  \\
\hline
J0212$+$5222$^{\dag,\ddag}$ & 02:12:52.2(6) & $+$52:22:45(13) & 376.386292(1) & 6.6(3) & 38 & 17 & 56373 - 56477 & 73\\
J0324$+$5239 & 03:24:55.46(4) & $+$52:39:31.3(2) & 336.620230291(3) & 0.381(3) &   119     &   64 & 55977 - 56478 & 309\\
J0426$+$4933 & 04:26:06.813(1) & $+$49:33:38.46(4) & 922.474730055(2) & 39.3444(2) &  88  &   101 & 55844 - 56478 & 240\\ 
J0555$+$3948$^{\ddag}$ & 05:55(5) & $+$39:48(5)  & 1146.9058(2) &  & 37 & 12 & 56373 - 56475 & 2904\\
J1905$-$0056$^{\ddag}$  & 19:05:27.9(1) & $-$00:56:37(5) & 214.3943414(3) & 1.07(7) & 227 & 13 & 56356 - 56482 & 115 \\
J1913$+$3732 & 19:13:27.887(3) & $+$37:32:12.30(7) & 851.078948902(7) & 1.3792(7) &   69    &  77 &  55977 - 56479 & 321\\ 
J1946$+$3417 & 19:46:25.13182(6) & $+$34:17:14.677(1) & 3.170139227806(2)  &  0.0000037(2) &  110  &   156 & 56089 - 56477 & 5 \\
J1959$+$3620 & 19:59:38.03(2) & $+$36:20:29.1(3) & 406.08118100(1) & 0.036(1) &  273   &        116 & 55839 - 56481 & 3069 \\
J2004$+$3429 & 20:04:46.97(3) & $+$34:29:17.7(5) & 240.95264193(1) & 206.825(4) &   351    &   90 & 56069 - 56482 & 3476\\ 
J2005$+$3552 & 20:05:47.50(6) & $+$35:52:24.3(1) & 307.94290464(2) & 2.99(1) &      445     &  55 & 56093 - 56480 & 512 \\
J2036$+$2835 & 20:36:46.363(5) & $+$28:35:10.44(7) & 1358.72676315(2) & 2.090(2) &  99   &     97 & 55873 - 56480 & 509\\ 
J2206$+$6151 & 22:06:18.119(6) & $+$61:51:58.10(3) & 322.673549948(5) & 0.397(4) &   167     &  47 & 56028 - 56478 & 230 \\
J2216$+$5759 & 22:16:05.22(3) & $+$57:59:53.7(3) & 419.10226464(2) & 69.048(2) &    176    & 97 & 55838 - 56478 & 3710\\ 
J2319$+$6411 & 23:19:35.210(2) & $+$64:11:25.755(7) & 216.01827884014(6) & 0.1632(5) &  246  &    56 & 56001 - 56475 & 75\\
J2333$+$6145 & 23:33:19.448(5) & $+$61:45:30.09(3) & 756.899382059(7) & 1.1761(6) &     125   & 94 & 55838 - 56475 & 412\\
\hline
\multicolumn{5}{l}{$^{\dag}$ co-discovered with the GBNCC survey \citep{Lynch2013}.}\\
\multicolumn{5}{l}{$^{\ddag}$ parameters measured w.r.t. reference epoch MJD 56375.}\\
\end{tabular}
\end{table*}

\begin{table*}
\centering
\caption{Further parameters for the first discoveries of the HTRU-North survey. Here, $D_{\rm DM}$ is the DM-derived distance, $B_{\rm surf}$ is the characteristic surface magnetic field strength, $\tau_{\rm c}$ is the characteristic age, $\dot{E}$ is the spin-down luminosity and $l$ and $b$ are the Galactic longitude and latitude, respectively. All DM-derived distances were calculated using the NE2001 Galactic free electron density model \citep{Cordes2002}, giving a likely uncertainty of $\gtrsim 20\%$ \citep{Deller2009}. \label{htrutab:derived}}
\begin{tabular}{lllllllll}
\hline
PSR & $l$ & $b$ &  $D_{\rm{DM}}$ & Mean flux density & $\log_{10}(\tau_{\rm{c}})$ & $\log_{10}(B_{\rm{surf}})$ & $\log_{10}(\dot{E})$\\
       & ($^{\circ}$) & ($^{\circ}$) &   (kpc)                    & at 1.5 GHz (mJy)   &  (years)      &  (Gauss)                       & (ergs s$^{-1}$)\\
\hline
J0212$+$5222 & 135.33 & $-$8.52 & 1.5 & 0.9(4) & 5.91 & 12.18 & 33.83 \\
J0324$+$5239 &  145.09  &  $-$3.48 &  3.6 & 0.19(4) & 7.15  &  11.56  &  32.60 \\
J0426$+$4933 &  154.44  &  0.29 &  2.4 & 0.19(5) &  5.56  &  12.79  &  33.30 \\
J0555$+$3948 & 171.62 & 7.21 & 1.2 & $\geq$0.09 \\
J1905$-$0056  & 33.69 & $-$3.55 & 5.9 & 0.11(1) & 6.50 & 11.69 & 33.63 \\
J1913$+$3732 &  69.10  &  12.13 &  4.0 & 0.38(5) &  6.99 &  12.04  &  31.94 \\
J1946$+$3417 &  69.29  &  4.71  &  5.1 & 0.29(6) & 10.14 &  8.03 & 33.65\\ 
J1959$+$3620 &  72.44  &  3.44 &   11.5 & 0.4(1) & 8.25  &  11.09  &  31.33\\
J2004$+$3429 &  71.42  &  1.57  &   12.5 & 0.11(4) & 4.26  &  12.85  &  35.76 \\
J2005$+$3552 &  72.71  &  2.14 &   $>$ 18.0 & 0.21(7) & 6.21  &  11.99  &  33.61\\
J2036$+$2835 &  70.41  &  $-$7.38 &   5.0 & 0.15(6) & 7.02  &  12.22  &  31.51 \\
J2206$+$6151  &  104.73  &  4.98 &  6.6 & 0.8(2) &  7.16  &  11.53  &  32.62 \\
J2216$+$5759 &  103.52  &  1.11 &   5.6 & 0.23(6) & 4.98  &  12.73  &  34.57\\
J2319$+$6411 &  113.14  &  3.08 &   $>$ 12.3  & 0.27(7) & 7.35  &  11.26  &  32.77\\
J2333$+$6145  &  113.83  &  0.27 &   5.3 & 0.47(7) & 7.00  &  11.98  &  32.03\\
\hline
\end{tabular}
\end{table*}

\subsection{Orion-spur observations}
\label{htru:oricyg}
To achieve complete coverage and statistics for a sample portion of the survey, mid-latitude pointings were targeted on the region $64.1^{\circ} < l < 71.9^{\circ}$, $|b|<15^{\circ}$. In this
direction the line-of-sight lies along the axis of the Orion spur up
to a distance of $\sim3$ kpc, and intersects with the Perseus arm and
Outer arm at distances of $\sim$ 6 kpc and $\sim 10$ kpc, respectively. These hydrogen-rich regions are known for star formation,
and as such make excellent targets for pulsar searches.

The only survey of comparable sensitivity to have covered this area
is the on-going P-ALFA survey \citep{Cordes2006}. Although the P-ALFA survey
discovered five new pulsars in the region, the
declination limit of $+38^{\circ}$ imposed by the Arecibo telescope
limited its coverage. The Effelsberg telescope has no
upper declination limit and so is capable of observing the entire
Orion-spur region.

Data were processed in both quick-look and full pipelines and all
candidates with folded profile significance greater than 6$\sigma$ were
viewed by eye. This has resulted in the discovery of six previously
unknown radio pulsars including the eccentric binary MSP~J1946$+$3417 and young pulsar J2004$+$3429, which are discussed in more detail in the subsequent sections.
\begin{figure}
\centering
\includegraphics[keepaspectratio=true,scale=0.45]{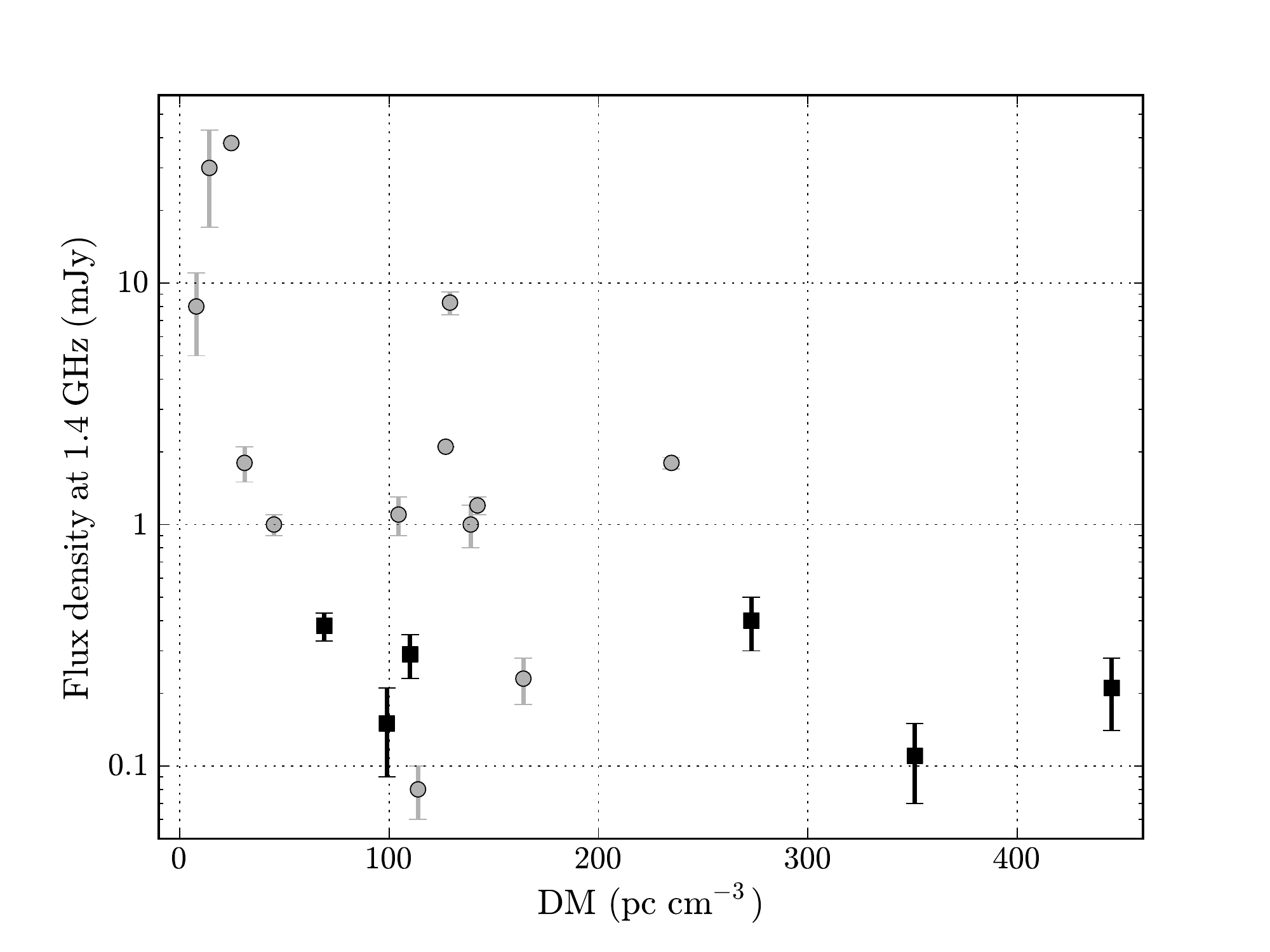}
\caption[Comparison of newly discovered pulsar population to existing pulsar population in the Orion-spur region.]{Comparison of the known pulsar population in the Orion-spur region, to pulsars discovered in this work. Grey circles show previously known pulsars with measured flux densities at 1.4 GHz. Black squares show pulsars discovered in this work.\label{htrufig:popcomp}}
\end{figure}

Figure \ref{htrufig:popcomp} shows the comparison between the newly discovered pulsars from this work and the known population in the region with measured flux densities at 1.4 GHz. It is clear that the pulsars discovered in this work have considerably lower flux densities at 1.4 GHz than their known counterparts. We also note that three of the newly discovered pulsars have DMs greater than 250 pc cm$^{-3}$. The only other blind survey to have found pulsars in this region at such high DMs is the P-ALFA survey, which stresses the importance of using backends with high frequency resolution. 

Of the high-DM pulsars discovered in the survey, PSRs J2005$+$3552 and J2319$+$6411 have DMs that are larger than the NE2001 Galactic free electron density model's \citep{Cordes2002} expected maximum DM contribution along their respective lines of sight. This results in DM-derived distances that place the pulsars outside of the Galaxy. However, in both cases the DM excess in the line of sight (75 and 13 pc cm$^{-3}$ for PSRs J2005$+$3552 and J2319$+$6411, respectively) falls within the expected uncertainties for the NE2001 model \citep{Deller2009}.

\subsection{PSR~1946$+$3417}
\label{htur:1946}
Discovered as a 7-$\sigma$ candidate in the full processing pipeline, PSR~J1946$+$3417 has the honour of being the first millisecond pulsar discovered in the HTRU-North survey. With a spin period of 3.14 ms, binary period of 27 days and minimum companion mass of 0.21~M$_{\astrosun}$ , the pulsar would at first glance appear to be normal MSP binary with a likely white dwarf companion. However,   timing of the pulsar has shown the system to have a large eccentricity ($e\approx0.14$). Assuming that the pulsar was spun-up through the accretion of mass from the current companion, one would expect the orbit to have been strongly circularised through tidal friction. Therefore, the high eccentricity of PSR~J1946$+$3417 is somewhat anomalous. A possible explanation for the orbital configuration may be found in PSR J1903$+$0327 \citep{Champion2008}, a similar system which may have been formed in a hierarchical triple \citep{Freire2011}. An in-depth discussion of the possible origins of PSR~J1946$+$3417 will be presented in Barr~et~al.~(in~prep.).

\subsection{PSR~J2004$+$3429}
\label{htru:2004}
PSR~J2004$+$3429 was initially discovered as an 11-$\sigma$ candidate
in a 3-minute pointing analysed with the quick-look pipeline. The
pulsar has a spin period of 241 ms and a DM of 352 pc cm$^{-3}$. The radio profile of {PSR~J2004$+$3429} shows two components separated by 180$^{\circ}$ (see Figure~\ref{htrufig:profs}), suggesting that the pulsar may be an orthogonal rotator.
Using the NE2001 Galactic electron density model \citep{Cordes2002} we
estimate a distance to the pulsar of $\sim12.5$
kpc. Analysis of pulsar distance measures by \citet{Verbiest2012}
suggests that this distance is likely overestimated, with the real
distance lying closer to 10 kpc. Timing with the Lovell radio
telescope has led to a phase-coherent timing solution for
{PSR~J2004$+$3429} that shows it to have a large period derivative of
$2\times10^{-13}$. For spin-down caused purely by the emission of magnetic dipole radiation, we find the pulsar to have a characteristic age of $\sim18$ kyr.

\subsubsection{SNR association}

Considering the young age of {PSR~J2004$+$3429}, it is highly likely that the pulsar is associated with a SNR. Using the SIMBAD astronomical data archive\footnote{http://simbad.u-strasbg.fr/simbad/}, we find
three published SNRs within a 2-degree radius of {PSR~J2004$+$3429}:
SNRs G069.7$+$01.0 \citep{Kothes2006},
G070.7$+$01.2.\citep{Kulkarni1992,Cameron2007} and G069.4$+$01.2
\citep{Yoshita2000}. The field also contains the candidate SNR G070.0$+$02.0
\citep{Mavromatakis2009}. 

Small distance estimates to SNRs G070.0$+$02.0 \citep[$< 1$ kpc,]
[]{Mavromatakis2009}, G069.4$+$01.2 \citep[$\sim2.5$ kpc,][]{Yoshita2000} and G070.7$+$01.2 \citep[$4.5\pm1$ kpc,][]{Bally1989}
appear to rule out any possible association with
{PSR~J2004$+$3429}. Using a revised $\Sigma$-D relation, which relates the surface brightness and diameter of SNRs to their distances,
\citet{Case1998} estimated a distance of $\sim14.4$ kpc to SNR
G069.7$+$01.0. This should be treated with care, as distance estimates via the $\Sigma$-D relation have an average uncertainty of $\sim40\%$, and distance measurements to individual objects, especially those at large distances where population statistics are limited, may have much larger uncertainties \citep{Case1998}. However, the large inferred distance is consistent with both the DM distance calculated for {PSR~J2004$+$3429} and the findings of \citet{Kothes2006}, who note that {SNR~G069.7$+$01.0} is unpolarised at
1.42 GHz, possibly due to a large distance resulting in high beam depolarisation.

We calculate the separation of {PSR~J2004$+$3429} and {SNR~G069.7$+$01.0}
to be $360\pm^{131}_{82}$ pc, assuming that {SNR~G069.7$+$01.0} is
located at the DM distance of {PSR~J2004$+$3429}. For {PSR~J2004$+$3429}
to have originated in {SNR~G069.7$+$01.0} would therefore require, assuming that the characteristic age is a close approximation of the true age, that the pulsar has a transverse velocity of the order 10$^4$ km
s$^{-1}$. Using the model for the pulsar velocity distribution as found by \citet{Hobbs2005}, we find the probability of
{PSR~J2004$+$3429} having such a high transverse velocity to be
consistent with zero. The large distance to {PSR~J2004$+$3429} makes any measurement of the pulsar's proper motion highly unlikely. Unless {PSR~J2004$+$3429} is shown to have a large transverse velocity, we rule out any association with {SNR~G069.7$+$01.0}.

\subsubsection{Gamma-ray observations}

{PSR~J2004$+$3429's} high spin-down luminosity ($\dot{E} = 5.7\times10^{35}$ ergs s$^{-1}$) marks the pulsar as a potential gamma-ray emitter \citep{2FGLcat}. To search for gamma-ray pulsations from {PSR~J2004$+$3429}, \emph{Fermi} LAT  photons recorded between 2008 August 4 and 2013 May 10, with energies above 0.1 GeV, and from a 3$^{\circ}$ region of interest interest (ROI) around {PSR~J2004$+$3429} were phase-folded using the ephemeris shown in {Table~\ref{htrutab:eph}} and the \emph{Fermi} plug-in distributed with the \textsc{tempo2} package \citep{Ray2011a}. To improve the chance of a detection, the phase-folded LAT data were restricted to `Source' class events of the P7\_V6 instrumental response functions. Furthermore, data taken during times when the rocking angle of the LAT exceed 52$^{\circ}$ or the Earth's limb infringed on the ROI were rejected. 

To determine if a statistically significant signal was present, a range of different angular and energy cuts was applied to the data in order to optimise the H-test parameter \citep{DeJager2010}. We tried maximum angular separation values between 0.1 and 3$^\circ$, and minimum photon energies ranging from 0.1 to 1 GeV. None of the cuts applied resulted in a greater than 3$\sigma$ significance. To remove any error introduced by uncertainties in our ephemeris, the same procedure was repeated using only data which was taken during the validity interval of our ephemeris. Again, no signal of greater than 3$\sigma$ significance was found. Considering the large distance to the pulsar, it is feasible that the pulsar will become detectable when more LAT data are available.

\section{Conclusion}
\label{htru:conclusion}
We have described the instrumentation, observing strategy,
sensitivity and expected results of the High Time-Resolution Universe
North (HTRU-North) pulsar survey, the first major search for radio
pulsars conducted with the 100-m Effelsberg radio telescope and
the most sensitive survey ever to observe the entire region above
$+30^{\circ}$ declination. The survey has thus far resulted in the
discovery of 15 radio pulsars, of which 13 have been found above
$+30^{\circ}$ declination.

Of the newly discovered pulsars, two are of particular
note. {PSR~1946$+$3417} is a highly eccentric MSP binary located in the
Galactic field. This system will be presented in detail in a future paper. {PSR~J2004$+$3429} is a young pulsar with a characteristic age
of $\sim$18 kyr. We currently rule out that {PSR~J2004$+$3429} is associated
with near-by {SNR~G069.7$+$01.0}, due to the high transverse velocity
required to place the pulsar at its current position with respect to
the remnant in the time scale suggested by its characteristic age. Despite its high spin-down luminosity, we find no evidence of gamma-ray emission from  {PSR~J2004$+$3429}. This is most likely due to the large distance to the pulsar.

\section*{Acknowledgements}
This work was carried out based on observations with the 100-m telescope of the
MPIfR (Max-Planck-Institut f\"{u}r Radioastronomie) at Effelsberg.

Pulsar research and observations at Jodrell Bank Observatory have been supported through Rolling Grants from the UK Science and Technology Facilities Council (STFC).

JPWV acknowledges support by the European Union under Marie-Curie
Intra-European Fellowship 236394.

PCCF and JPWV acknowledge support by the European Research Council
under ERC Starting Grant Beacon (contract no. 279702).

DLR and MAM acknowledge support from WVEPSCoR and the Research Corporation for Scientific Advancement.

KJL acknowledges support from the ERC Advanced Grant ``LEAP'', Grant Agreement Number 227947

\bibliographystyle{mnras}
\bibliography{library.bib}
\end{document}